
\input harvmac
\def\nopage{\footline={\hfil}}
\nopage
%
%
%
%
\def\vect#1#2{\setbox0=\hbox{$#1$}   \def\cpy{\copy0\kern-\wd0}
              \def\column{\raise.025em\cpy \raise.0125em\cpy \cpy}
              \def\back{\kern-#2em}  \def\forw{\kern+#2em}
        \hbox{\raise.0250em\cpy      \back\back\column
              \forw\column           \forw\forw\column
              \forw\column           \back\back\box0  } }
\edef\tfontsize{ scaled\magstep3}
 \tfontsize  \tfontsize
 \tfontsize \font\titlei=cmmi10 \tfontsize
\font\titleis=cmmi7 \tfontsize \font\titleiss=cmmi5 \tfontsize
\font\titlesy=cmsy10 \tfontsize \font\titlesys=cmsy7 \tfontsize
\font\titlesyss=cmsy5 \tfontsize  \tfontsize
\skewchar\titlei='177 \skewchar\titleis='177 \skewchar\titleiss='177
\skewchar\titlesy='60 \skewchar\titlesys='60 \skewchar\titlesyss='60
%
\global\newcount\meqno
\def\eqn#1#2{\xdef #1{(\secsym\the\meqno)}
\global\advance\meqno by 1 $$#2\eqno#1$$}
%
\global\newcount\refno\def\ref#1#2#3{\global\advance\refno by1
\xdef #1{[\the\refno]}\xdef #2{#3}#1}
%

%
\def\uplrarrow#1{\raise1.5ex\hbox{$\leftrightarrow$}\mkern-16.5mu #1}
%
%
%
\def\ignore#1{}
\def\pair{e^+ e^-}
\def\bbar{\bar B}
\def\u{\Upsilon}
\def\two{{B\bbar}}

\def\bbpi{{BB\pi}}
\def\three{{B B^\pm \pi^\mp}}

\def\gev{\,{\rm GeV}}

%

\def\ub{\underline{b}}

%
%

%

%

%
\def\bB{{\bf B}}
\def\oB{\overline{\bB}}
\def\uB{{\bf B}\hskip-.715em{\underline{\hphantom n}}\hskip.15em}
\def\ouB{\overline{\uB}}

\def\fpi{f_\pi}
%

\def\l{{\cal L}}
\def\leff{\l_{\rm eff}}
\def\tr{{\rm tr}}


\def\dk{\Delta k}
\def\k{\vect{k}{0.0125}}

\def\dkb{\Delta\k}

\def\epi{E_\pi}

\def\ppi{p_\pi}
\def\mpi{m_\pi}

\def\dm{\Delta M}
\def\pp{{\rm PP}}
\def\pv{{\rm PV}}
\def\vv{{\rm VV}}

\def\Rpp{R_\pp}
\def\Rpv{R_\pv}
\def\Rvv{R_\vv}
\def\Rtwo{R_{2\pi}}
\def\so{\sigma_0}

\def\crsc#1{\sigma(\pair\to #1)}
\def\rts{\sqrt{s}}

\def\ak{\overline{k}}
\def\akb{\overline{\k}}
\def\Ra{R_\alpha}
\def\Rai{\Ra^i}

\def\ra{r_\alpha}



%
\def\lapprox{\mathrel{\hbox{\setbox0=\hbox{$\sim$}
\lower.8ex\copy0\kern-\wd0\raise.3ex\hbox{$<$}}}}
\def\gapprox{\mathrel{\hbox{\setbox0=\hbox{$\sim$}
\lower.8ex\copy0\kern-\wd0\raise.3ex\hbox{$>$}}}}
%

\hsize 17truecm
\hoffset -0.3truecm
\vsize 24.94truecm
\voffset -1.25truecm
%

%
%
%
\nopage
\phantom{a}
\vskip 4.85truecm
\centerline{{ THE RATE FOR $B\bbar$ PRODUCTION
ACCOMPANIED BY A SINGLE PION}}
\vskip 0.75truecm
{\baselineskip=0.45truecm
\centerline{Lisa Randall and Eric Sather}
\centerline{Center for Theoretical Physics, MIT,
Cambridge, MA 02139,  USA}
\vfill}
\vskip -2.1truecm
\noindent
ABSTRACT
\vskip 0.55truecm
{\baselineskip=0.45truecm
\noindent
We study the rate for the production of $\three$, where
the sign of the charged pion tags the flavor content
of the neutral $B$ meson. We estimate this
branching ratio, employing the heavy meson
chiral effective theory. We find that at center of mass
energy of approximately 12 GeV, a $B$ meson pair
should be produced as often with and without an
accompanying charged pion.  We also
calculate two pion production at this center of mass
energy, and find that it is negligible, as is the
rate for rho production. We consider
the implications for CP violation studies.
\vfill}
{\vbox{\hbox{MIT-CTP
2214}}}{May 1993}
\eject

\pageno=1
\nopage
\baselineskip 24pt plus 2pt minus 2pt
\xdef\secsym{}\global\meqno=1
\medskip
We study an alternative method for tagging neutral $B$
mesons for the purpose of CP violation studies which
can be used at a symmetric collider.
This talk is based on ref. ~\ref\lrs\lrsref{Laurent Lellouch, Lisa Randall and
Eric Sather,
MIT-CTP 2155 (1992).}We consider Yamamoto's   suggestion
\ref\yam\yamref{H.~Yamamoto, Cornell University note CBX 92-94, to be
published.}\
to study  $B$ mesons produced
in conjunction with a single charged pion at or near an $\u$ resonance, so
that the  sign of the charged pion tags the single neutral $B$ meson
as $B^0$ or $\bbar^0$.  Neutral $B^*$ mesons can also be used, since
they decay immediately via photon emission into pseudoscalars,
before weak mixing or decays have occurred. This method should
provide a simple and efficient tag of the neutral $B$ meson: in
addition to the decay products of the neutral $B$, one only needs to
detect the additional soft pion. The charged $B$ is then tagged by
the invariant mass of the missing four momentum, so it need not be
reconstructed.  Unlike many conventional proposals for the study  of
CP violation in the neutral $B$ system, essentially all the events
are tagged.

The utility of this method depends on the event rate. We calculate
this rate, employing the heavy meson chiral effective theory, in
order to evaluate the potential of Yamamoto's proposal.
We find that Yamamoto's proposal
could prove  a competitive method for tagging neutral $B$'s for the
study of CP violation at a symmetric collider at slightly higher
center of mass energy, about $12\gev$, where nevertheless multipion final
states still have a negligible branching fraction.

We first construct the effective lagrangian for $BB$
and $\bbpi$ production. Since the kinetic
energies of the $B$ and $\bbar$ are on the order of the QCD scale,
both the $B$ and $\bbar$ mesons have essentially timelike four
velocities in the center of mass frame: $v^\mu\approx
v'^\mu\approx(1,0,0,0)$. We couple a  source $S^\mu$  to the
$b$-quark current as  $\overline{b}_{v'}S^\mu\gamma_\mu\overline{\ub}_v$
(i.e., with the heavy quark spins coupled to the spin of the
source). When we match onto the heavy meson theory, we can couple
the source to the mesons as $\oB(v')S^\mu\gamma_\mu\ouB(v)$. This
gives the same matrix elements as if we had coupled a source to
heavy quarks, and then evaluated the heavy quark matrix elements.

The lagrangian  applicable to low-energy production of $B$ and
$\bbar$ meson is
\eqn\efflag{\eqalign{
\leff=&-i\tr\{\oB_a(v)v^\mu\partial_\mu\bB_a(v)\}
-i\tr\{\ouB_a(v)v^\mu\partial_\mu\uB_a(v)\}\cr
&+g\tr\{\oB_a(v)\bB_b(v) A^\nu_{ba}\gamma_\nu\gamma_5\}
+g\tr\{\uB_a(v)\ouB_b(v) A^\nu_{ba}\gamma_\nu\gamma_5\}\cr
&+\l_S,\cr
\l_S=&{-i\lambda\over2}S^\mu\tr\{\gamma_\mu\ouB_a(v)
\uplrarrow{D}_{ab}^\nu\gamma_\nu\oB_b(v)\}\cr
&+\lambda g' S^\mu\tr\{\gamma_\mu\ouB_a(v)
A^\nu_{ab}\gamma_\nu\gamma_5\oB_b(v)\}.}}
Here $D=\partial+V$ is the chiral covariant derivative incorporating
the pion fields. As usual, a factor of $\sqrt{M_B}$ has been
absorbed into the heavy meson fields along with the
position-dependent phase  corresponding to the momentum of the heavy
quark (so that a derivative acting on these fields only gives a
factor of the residual momentum), in order to suppress the
appearance of the heavy quark mass and emphasize the heavy quark
symmetry. Because this is the low energy theory, no large momentum
transfers are permitted. At higher energies, the appropriate
lagrangian would be the heavy meson lagrangian with velocities $v
\ne v'$. The result for two meson production matches smoothly, as
the difference in residual momenta in the amplitude gets replaced by
the difference in heavy meson velocities.

The kinetic and axial coupling terms for the $B$ mesons have been
discussed previously and result from the straightforward application
of heavy quark and chiral effective field theories \ref\wse\wseref{M.~B.~Wise,
Phys.\ Rev.\ D 45 (1992) 2188.}. ${\cal
L}_S$ is the new term and follows from the assumptions  described
above.   Note that with the trace the heavy quark spin labels are
coupled to $S^\mu\gamma_\mu$.

We now summarize the result of calculating the process of interest.
{}From the lagrangian ~\efflag\ we see that two types of diagrams
contribute to $\bbpi$ production. The pion can be produced
``indirectly'' by being emitted from a virtual $B$ meson through the
heavy-meson axial coupling. Or, the pion can be produced
``directly'', together with the $B$ mesons at a single vertex. This
diagram comes from  the contact term in the lagrangian in  which the
source couples directly to the $B$ meson and axial fields.  The direct
contribution is much the smaller of the two,
because it is higher order in the heavy quark
mass expansion \lrs,
so the discussion concentrates on the indirect contribution.

In order to compare $\three$ with $\two$ as sources of neutral   $B$
mesons we normalize the $\three$ cross sections by dividing them
by the cross section for $\pair\to$ neutral $B$ mesons, $\so$.
The density of states for a final state of $B$ mesons and a pion
simplifies for nonrelativistic $B$ mesons to
\eqn\dos{\eqalign{D=&{1\over64\pi^3}M_B|\dkb|\,|\akb|\,d\epi\,
d(\cos\theta)\cr
=&{1\over64\pi^3}M_B^{3/2}\ppi(r-\epi)^{1/2}d\epi\,d(\cos\theta).}}
where $r$ is the center of mass energy that remains after supplying
the rest mass energies of the heavy mesons.
As discussed in the last section, we treat the $B$ mesons as
nonrelativistic in the laboratory frame, working to lowest
nonvanishing order in the heavy-meson three-momenta. Implicit in our
approximations is the realization that the kinetic energy is fairly
evenly shared between the pion and the heavy mesons.Therefore, when we work at
energies where the heavy mesons are
nonrelativistic, there is a hierarchy of energy scales,
\eqn\scales{T_B,\ \epi,\ \ppi\ (\propto M_B^0)\ \ll |\k_B|
\ (\propto M_B^{1/2})\ll\ M_B,}
where $T_B$ is the kinetic energy of the $B$ mesons.  Since the
dimensionful quantities that compensate the different powers of
$M_B$ here are of order $r$, we are essentially working to lowest
order in $r/M_B$.  Taking sums and differences of the $B$ meson
momenta, this hierarchy can be reexpressed as
\eqn\adscales{\ak^0,\dk^0,|\akb|\ (\propto M_B^0)\ \ll|\dkb|
\ (\propto M_B^{1/2})\ll\ M_B.}
In calculating a given amplitude, we retain only the leading term
according to this hierarchy. Specifically, we drop  $\ak^0$ and
$\dk^0$ compared to $M_B$, incurring errors of order $r/M_B$.  We
also drop $|\akb|$ compared to $|\dkb|$ and $|\dkb|$ compared to
$M_B$.  This would seem to mean dropping terms of order
$\sqrt{r/M_B}$. However, once an amplitude is squared,
averaged/summed over initial/final polarizations, and integrated
over $\cos\theta$, the result depends only on $\akb^2$ and $\dkb^2$
($\akb\cdot\dkb$ vanishes in the angular integration).  Hence all
the dropped terms  are smaller by a factor of $r/M_B$.
The result of calculating the
cross-section ratios for the various $\three$ final states  is
\eqn\ratioresults{
\Rai(s)={2g^2\over 3\pi^2\fpi^2}{1\over P(s)}
\int_{\mpi}^{\ra} d\epi \ppi^3 (\ra-\epi)^{3/2}
\times\cases{{1\over(\epi-\dm)^2},&$\alpha$=PP;\cr
\noalign{\vskip4pt}
{7/4\over(\epi-\dm)^2}+{1/2\over\epi^2-(\dm)^2}\cr
\quad\quad\,\,+{1\over\epi^2}+{3/4\over(\epi+\dm)^2},&$\alpha$=PV;\cr
\noalign{\vskip4pt}
{5\over\epi^2}+{2\over(\epi+\dm)^2},&$\alpha$=VV.\cr} }
Here the $i$ signifies that these are the indirect contributions to
$\Ra$. The upper integration limit is the value of $r$ appropriate
to the heavy-meson content of the $\three$ final state.
\def\sp{$\ $}
\def\row#1#2#3#4#5#6#7#8#9
      {&$#1$&&$#2$&&$#3$&&$#4$&&$#5$&&$#6$&&$#7$&&$#8$&&$#9$&\cr}
\def\rowomit{height2pt
      &\omit&&\omit&&\omit&&\omit&&\omit&&\omit&&\omit&&\omit&&\omit&\cr}
\def\horizrule{\rowomit\noalign{\hrule}\rowomit}
\topinsert
\centerline{Table 1. The ratios
$\crsc{\three}/\crsc{\two}$ expressed as percentages.}
$$\vbox{\offinterlineskip
\hrule\halign{
\vrule#&\sp\hfil#\hfil\sp& \vrule#&\sp\hfil#\hfil\sp&
\vrule#&\sp\hfil#\hfil\sp& \vrule#&\sp\hfil#\hfil\sp&
\vrule#&\sp\hfil#\hfil\sp& \vrule#&\sp\hfil#\hfil\sp&
\vrule#&\sp\hfil#\hfil\sp& \vrule#&\sp\hfil#\hfil\sp&
\vrule#&\sp\strut\hfil#\hfil\sp&\vrule#\cr
\rowomit
\row{\sqrt{s}({\rm MeV})}{\Rpp^i}{\Rpv^i}{\Rvv^i}{R^i}{\Rpv^d}{\Rvv^d
}{R^d}{R}
\horizrule
 \row{10865}{0.12}{0.12}{0.02}{0.26}{0.03}{0.01}{0.04}{0.30}
 \row{11020}{0.53}{0.99}{0.70}{2.2 }{0.19}{0.10}{0.29}{2.5 }
 \row{11200}{1.3 }{3.1 }{3.1 }{7.6 }{0.59}{0.40}{0.99}{8.6 }
 \row{11500}{3.3 }{9.2 }{11. }{24. }{2.0 }{1.6 }{3.7 }{27. }
 \row{12000}{8.2 }{26. }{36. }{70. }{7.8 }{6.7 }{14. }{85. }
\rowomit
\noalign{\hrule}
}}$$
\endinsert

The axial coupling of the heavy mesons,  $g$, has been bounded above
by $g^2\le0.5$ \ref\amd\amdref{J.~Amundson, C.~G.~Boyd, E.~Jenkins,
M.~Luke, A.~Manohar, J.~Rosner, M.~Savage, M.~Wise,  Phys. Lett. ~B296
(1992).} using the experimental upper limit for the
$D^{*+}$ width \ref\acc\accref{ACCMOR Collaboration. S.~Barlag, et
al.\ Phys.\ Lett.\ B278 (1992) 480.}  which is dominated by
$D^{*+}\to D^0\pi^+$ and $D^{*+}\to D^+\pi^0$. Using the maximum
allowed value of $g^2$ and taking $g'^2=1$,   the numerical values
of the ratios $\Ra$ (expressed as percentages) are given in Table~1
for various values of the center-of-mass energy, $\sqrt{s}$,
where we have included both indirect
and direct contributions. We see that although production in the
 resonance regime is suppressed,  for center of mass
energy of $12\gev$, the rate is almost comparable to the rate
without a pion.

We supplement this calculation  with a calculation of
two accompanying pions. We find this rate is very small
in comparison with the single pion rate calculated above. We
briefly outline this calculation.

We first consider a general vector coupling of the $B$'s which
could produce either a rho meson or a pair of pions. It is
straightforward to show that there is a cancellation between
the vector emission from the $B$ and the $\bbar$ states.
The coupling of the $B$ mesons to the vector field $V$ has the
form
\eqn\vector{\eqalign{\l_{B\bbar V}
=&-i\tr\{\oB_a(v)\bB_b(v)V_{ba}^\mu\gamma_\mu \}
-i\tr\{\ouB_a(v)\uB_b(v) V_{ba}^\mu\gamma_\mu \}\cr
=&i(v\cdot V_{ba})(\tr\{\oB_a(v)\bB_b(v) \}
-\tr\{\ouB_a(v)\uB_b(v) \}).}}
The traces in the second line just give the normalization of the
heavy meson fields. Hence the Feynman rule for vector emission  is
independent of the heavy meson spin state, but has opposite signs for
$B$ and $\bbar$ mesons.   Then, because the propagator of the
intermediate heavy meson  is the same whether the vector state is
emitted from a $B$ or $\bbar$ meson, the amplitudes for these two
processes cancel.

Therefore, $\rho$ production is higher order in the heavy
quark expansion.
Moreover,  of the eight possible graphs which would contribute
to production of a two pion final state, only six contribute.  These are
readily calculated (here we take the heavy quark limit, neglecting
$\Delta M$) by summing over intermediate states and using the
completeness relations for the $B$ mesons fields.

The nonrelativistic four body phase space integrated over angles is
\eqn\fb{D_{2\pi}=
{1\over256\pi^5} M_B^{3/2} (r-E_1-E_2)^{1/2} p_1\,p_2\,dE_1\,dE_2}
where $E_i$ and $p_i$ are the energies and momenta of the
two pions.  After we have summed over all possible spin and isospin states
of the $B$'s and pions, we find that the total production rate of
$B$ mesons and two pions, normalized to $\so$, is given by
\eqn\rtwopi{\Rtwo={g^4\over24\pi^4\fpi^4}
\int_{\mpi}^{r-\mpi}\! dE_1 \int_{\mpi}^{r-E_1}\! dE_2\,
p_1^3\, p_2^3 \left({r-E_1-E_2\over r}\right)^{3/2}
{2(E_1^2+E_2^2)+3E_1E_2\over E_1^2 E_2^2(E_1+E_2)^2}\ .}
When we integrate over the full final phase space
we find $\Rtwo=4\%$ at $\rts=12\gev$ and $\Rtwo=17\%$ at
$\rts=12.5\gev$.
This indicates
that there should be some energy between the resonance regime and
high energy where single pion emission is not suppressed but multipion
final states are nonetheless small.
Note this is very different from what one would have
guessed by a naive application of Poisson statistics.

It is clear that the proposal of  Yamamoto is quite interesting.  It
appears that there might be a sufficiently large rate for
self--tagging $B$ meson events  for this to be a viable method of
studying CP violation for neutral mesons at a symmetric collider.
Despite the  limitations of our calculation, we can nevertheless
establish several interesting results.  Within the resonance regime,
the process we consider will probably not occur at a sufficiently
large rate to compete with the more conventional proposal for the
study of CP violation at a symmetric collider, namely $BB^*$
production followed by $B^*\to B\gamma$.   At center of mass energy
of about $12\gev$, pion emission from a $B$ meson pair could
occur as often as not. Moreover,
multipion production should be negligible.
 Although the calculation here is reliable
only  over about half the range of pion energies, it is clear that
the rate for a single pion accompanying the $B$ mesons is large,
even from this restricted phase space. Of course, a large number
of $B's$ is always required, which could prove difficult to attain.
Probably only for the larger possible angles will CP violation
prove accessible at a symmetric collider.

\baselineskip 12pt plus 1pt minus 1pt
\centerline{\bf References}
\bigskip
\item{\lrs}\lrsref
\medskip
\item{\yam}\yamref
\medskip
\medskip
\item{\wse}\wseref
\medskip
\item{\amd}\amdref
\medskip
\item{\acc}\accref
\medskip
\vfill
\eject
\end